\documentclass[final,5p,times,twocolumn]{elsarticle}

\usepackage{graphicx}

\usepackage{amsmath}
\usepackage{amssymb}

\newcommand{\xmds}{\textsc{XMDS2}}
\newcommand{\abs}[1]{\left|#1\right|}

\journal{Computer Physics Communications}

\begin{document}

\begin{frontmatter}



\title{\xmds: Fast, scalable simulation of coupled stochastic partial differential equations}


\author{Graham R. Dennis\corref{cor1}}
\cortext[cor1]{graham.dennis@anu.edu.au}
\author{Joseph J. Hope\corref{}}
\author{Mattias T. Johnsson\corref{}}

\address{Research School of Physics and Engineering, The Australian National University, Canberra, ACT 0200, Australia}

\begin{abstract}
\xmds\ is a cross-platform, GPL-licensed, open source package for numerically integrating initial value problems that range from a single ordinary differential equation up to systems of coupled stochastic partial differential equations.  The equations are described in a high-level XML-based script, and the package generates low-level optionally parallelised C++ code for the efficient solution of those equations.  It combines the advantages of high-level simulations, namely fast and low-error development, with the speed, portability and scalability of hand-written code.  \xmds\ is a complete redesign of the XMDS package, and features support for a much wider problem space while also producing faster code.  
\end{abstract}



\end{frontmatter}



\noindent
{\bf Program summary}

\begin{small}

\noindent
{\em Manuscript Title:} \xmds: Fast, scalable simulation of coupled stochastic partial differential equations                                       \\
{\em Authors:}  Graham R. Dennis, Joseph J. Hope, Mattias T. Johnsson                                              \\
{\em Program Title:}  XMDS2                                        \\
{\em Journal Reference:}                                      \\
{\em Catalogue identifier:}                                   \\
{\em Licensing provisions:} GNU Public License 2                                   \\
{\em Programming language:} Python and C++.                                   \\
{\em Computer:}  Any computer with a Unix-like system, a C++ compiler and Python                                             \\
{\em Operating system:} Any Unix-like system; developed under Mac OS X and GNU/Linux                                       \\
{\em RAM:} Problem dependent (roughly 50 bytes per grid point)                                              \\
{\em Number of processors used:} Up to the minimum of the number of points in each of the first two dimensions                             
{\em Classification:} 4.3 Differential Equations, 6.5 Software including Parallel Algorithms                                        \\
{\em External routines/libraries:} The external libraries required are problem-dependent. Uses FFTW3 Fourier transforms (used only for FFT-based spectral methods), dSFMT random number generation (used only for stochastic problems), MPI message-passing interface (used only for distributed problems), HDF5, GNU Scientific Library (used only for Bessel-based spectral methods) and a BLAS implementation (used only for non-FFT-based spectral methods).                           \\
{\em Nature of problem:}   General coupled initial-value stochastic partial differential equations   \\
{\em Solution method:} Spectral method with method-of-lines integration \\
{\em Running time:} Determined by the size of the problem \\
{\em Web site: } \verb+http://www.xmds.org+
\end{small}

\section{Introduction}
\label{sec:Introduction}
The integration of a system of variables from a set of initial conditions is one of the most widely performed tasks in quantitative simulation.  Numerical integration is typically performed in one of two different styles: high-level methods using general software tools, or low-level methods using bespoke hand-tuned source code.  The high-level approach requires much less code, and is therefore fast to develop and comparatively free of coding errors.  However, the low-level approach can provide dramatic and necessary performance improvements, can utilise the full capacity of the computing platform for which it is developed, and is more customisable.  \xmds\ is a software package whose aim is to provide the key benefits of both approaches \cite{XMDSWebsite}.

The purpose of \xmds\ is to simplify the process of creating simulations that solve systems of initial-value partial and ordinary differential equations. Instead of going through the error-prone process of hand-writing thousands of lines of code, \xmds\ enables problems to be described in a simple XML format. From this XML description \xmds\ generates a C++ simulation that solves the problem using fast algorithms. The code generated by \xmds\ is typically as fast as, or faster than, hand-written code, but by using \xmds\, the time taken to produce the simulation is significantly reduced.

\xmds\ can be used to simulate almost any set of (coupled) (partial) (stochastic) differential equations in any number of dimensions.  It can input and output data in a range of data formats, produce programs that can take command-line arguments, and produce parallelised code suitable for either modern computer architectures or distributed clusters.

Aside from innumerable low-level libraries and high-level packages for numerical integration, there have also been multiple previous attempts to automate or semi-automate the process of coding low-level numerical simulations, such as \cite{DeRose94anenvironment,RNPL}.  
Rather than provide `shell code' that can be edited, an \xmds\ script is effectively a self-contained language written in XML which is used to generate a fast C++ simulation.

The first version of XMDS was released in 1997 as an open-source software package written in C++ which could simulate a class of stochastic partial differential equations \citep{Collecutt:2001}.  Over the next decade, it was then expanded in scope and features by a growing group of developers.  While most of these developers came from the fields of quantum optics and atom optics, where its ability to integrate stochastic equations is particularly pertinent, the package slowly gained wider popularity.  In 2008, the decision was taken to completely rewrite the package in Python (although still generating low-level C++ code), with a re-engineered structure that would allow it to address a much wider problem space.  \xmds\ was released in 2010, and has recently received its first major update, along with extensive documentation, installers, and an examples library.

Citing only a few examples, XMDS and increasingly \xmds\ have been used in the fields of quantum atom optics \cite{Dall:2007,Dall:2009}, quantum optics \cite{Hsu:2006, Hetet:2008}, quantum control \cite{Szigeti:2010}, predator-prey dynamics \cite{Li:2012} and ecology \cite{Reichenbach:2007a, Reichenbach:2007b, Reichenbach:2008}.

\section{Problem class}

\xmds\ solves systems of initial-value differential equations.  Each differential equation can:
\begin{enumerate}
    \item have an arbitrary number of dimensions, which may differ from that of other differential equations in the system,
    \item involve integrals of quantities in the differential system, or
    \item include stochastic elements either in initial conditions and filters, or in the dynamical equations themselves.
\end{enumerate}

As an example, property~1 means that \xmds\ can solve systems in which a partial differential equation is coupled to an ordinary differential equation.  Property~2 allows the evolution of the ordinary differential equation to depend upon moments of the partial differential equation.  Property~2 also allows partial differential equations to depend non-locally on system quantities.  Property~3 permits the integration of systems of stochastic (partial) differential equations, which are typically written using either Gaussian noise (via a Wiener process), or a Poissonian noise in which the system changes state in a discontinuous way.

\begin{figure}
    \centering
    \includegraphics[width=8cm]{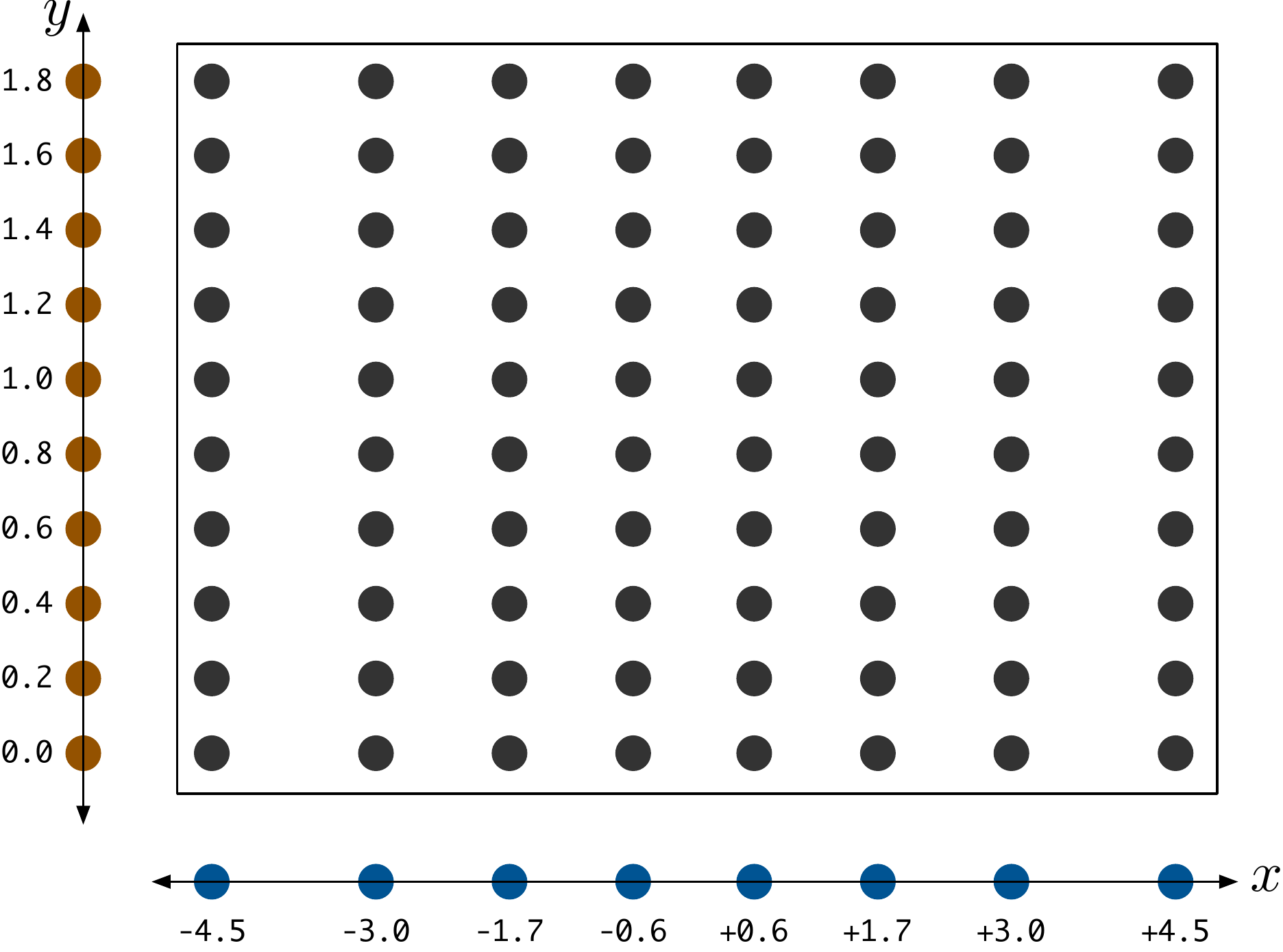}
    \caption{An example of a tensor product lattice. The specific unequal lattice spacing in the $x$ direction is due to the fact that the basis functions in that dimension have been chosen to be Hermite-Gauss.}
    \label{fig:TensorProductLattice}
\end{figure}

\xmds\ uses spectral methods \citep{SpectralMethods}, which induce two restrictions on the problem space.  The first is that the geometry of the simulation domain must be a tensor product of lattices in each dimension (see Figure~\ref{fig:TensorProductLattice}).  The second restriction is that the boundary conditions must be compatible with the spectral method used.  \xmds\ currently supports periodic, even, odd and zero boundary conditions.  Spectral methods allow the approximation of spatial derivatives with `exponential' accuracy (see Section~\ref{sec:Spectral}). In addition, the restriction to tensor product lattices affords significant computational savings which will be discussed later. The disadvantage is that \xmds\ cannot be used to solve problems on arbitrary-shaped domains as is possible using finite-element methods. This is not a significant limitation for a wide class of problems, as the system is often constrained to evolve within a finite domain. Quantum atom optics problems, for example, have trapping terms in the differential equation that cause the solution to be non-negligible over a finite domain. 

The use of spectral methods means that \xmds\ represents the solution as a linear combination of \emph{global} basis functions that extend over the entire domain.  This is an accurate representation for solutions which are smooth.  Problems which contain shocks or other spatial discontinuities (including discontinuous derivatives) are better served by \emph{local} methods such as finite difference or finite element methods.

Subject to these caveats, \xmds\ is applicable to a broad problem class and employs efficient and accurate algorithms for the solution of these problems.

\section{Algorithms employed}

\xmds\ employs efficient algorithms in its generated simulations.  These include:
\begin{enumerate}
    \item Spectral methods for computing spatial derivatives,
    \item Fast spatial-to-spectral transforms including FFTs and parity-exploiting matrix transforms,
    \item Distributed memory parallelism,
    \item Gaussian quadrature for spatial integration,
    \item Method-of-lines explicit temporal integration schemes, and
    \item Interaction picture methods for exactly solving linear parts of the problem.
\end{enumerate}

\subsection{The spectral method}
\label{sec:Spectral}

\xmds\ spatially discretises the problem by applying the spectral method \citep{SpectralMethods}. This method decomposes the solution as a weighted sum of a finite set of orthonormal basis functions.  For example, the quantity $f(x, y)$ is represented as
\begin{align}
	f(x, y) &= \sum_{n,m} F_{n,m} X_n(x) Y_m(y), \label{eq:SpectralDecomposition}
\end{align}
where $F_{n,m}$ is a matrix of coefficients, $X_n(x)$ is the $n^\text{th}$ basis function for the $x$ dimension, and $Y_m(y)$ is the $m^\text{th}$ basis function for the $y$ dimension.  The coefficients $F_{n,m}$ fully describe the solution and are the \emph{spectral} representation of the solution.  Typically in \xmds, the number of basis functions is equal to the number of grid points in each dimension.  In this case, the \emph{spectral} representation is equivalent to the \emph{spatial} representation $f(x_i, y_j)$, the values of the solution at the grid points.  The two representations are linked by the linear transformation \eqref{eq:SpectralDecomposition} and its inverse.  

Spectral methods approximate spatial derivatives using the decomposition \eqref{eq:SpectralDecomposition} and using analytic expressions for the derivatives of the basis functions,
\begin{align}
    \frac{\partial^p}{\partial x^p} \frac{\partial^q}{\partial y^q} f(x, y) &= \sum_{n,m} F_{n, m} \frac{d^p X_n(x)}{d x^p} \frac{d^q Y_m(y)}{d y^q}.
\end{align}

Spatial derivatives approximated in this manner are `exponentially' accurate.  In general, an optimal $M$-point method to calculate a $k$-order derivative of a function will have error $\mathcal{O}(h^{M-k})$ where $h$ is the grid-point spacing.  As $h\propto 1/N$ where $N$ is the number of grid points, such a method will converge like $\mathcal{O}(1/N^{M-k})$ for a $k$-order derivative.  In spectral methods the value of the solution at all grid points is used when computing spatial derivatives, hence $M=N$.  In this case $M$ increases as the number of grid points $N$ increases resulting in a method whose order effectively \emph{increases} as the number of grid points increases.  The asymptotic error of a spectral method is $\mathcal{O}(1/N^{N-k})$ which converges \emph{exponentially}.

The basis functions are typically chosen to make part of the differential equation diagonal in the spectral basis.  \xmds\ supports the following spectral methods for each dimension:
\begin{itemize}
	\item Fourier modes (complex exponentials),
	\begin{align*}
		X_n(x) &= e^{i k_n x}.
	\end{align*}
	This method imposes periodic boundary conditions. The basis functions are eigenfunctions of the Cartesian spatial derivative operator.  This is a general purpose method.
	\item Cosine / sine functions,
	\begin{align*}
		X_n(x) = \cos(k_n x) \quad \text{or}\quad X_n = \sin(k_n x).
	\end{align*}
	These methods impose even and odd boundary conditions respectively at the ends of the domain.  The basis functions are eigenfunctions of the Laplacian in Cartesian coordinates. This method is useful when the problem has even or odd reflection symmetry about a plane.
	\item `Cylindrical' Bessel functions,
	\begin{align*}
		R_n(r) = J_m(k_n r),
	\end{align*}
	where $J_m(r)$ is the order-$m$ Bessel function of the first kind. This method imposes analytic boundary conditions at the origin and zero Dirichlet boundary conditions at the outer boundary.  The basis functions are eigenfunctions of the radial component of the Laplacian in cylindrical coordinates.  This method is useful for problems with rotational symmetry. See \citep{Ronen:2006} for more details.
	\item `Spherical' Bessel functions,
	\begin{align*}
		R_n(r) = \sqrt{\frac{\pi}{2 r}}J_{l+\frac{1}{2}}(k_n r).
	\end{align*}
	This method imposes analytic boundary conditions at the origin and zero Dirichlet boundary conditions at the outer boundary.  The basis functions are eigenfunctions of the radial component of the Laplacian in spherical coordinates.  This method is useful for problems with spherical symmetry.
	\item Hermite-Gauss functions,
	\begin{align*}
		\psi_n(x) &= (2^n n! \sigma \sqrt{\pi})^{-1/2} e^{-x^2/2 \sigma^2} H_n(\sigma x),\quad \text{where:}\\
		H_n(x) &= (-1)^n e^{x^2} \frac{d^n}{dx^n} \left(e^{-x^2}\right)
	\end{align*}
	This method requires that the solution decay as $e^{-x^2/2\sigma^2}$ in the limit $x\rightarrow \pm \infty$.  The basis functions are eigenfunctions of the Schr\"o{}dinger equation for the harmonic oscillator:
	\begin{align}
		- \frac{\hbar^2}{2 m} \frac{\partial^2 \psi_n}{\partial x^2} + \frac{1}{2} m \omega^2 x^2 \psi_n(x) = \hbar \omega\left(n+\frac{1}{2}\right) \psi_n(x), \label{eq:HarmonicSchrodingerEquation}
	\end{align}
	with $\sigma = \sqrt{\hbar / (m \omega)}$.  This method is useful for solving problems similar to \eqref{eq:HarmonicSchrodingerEquation} with nonlinear terms.
\end{itemize}

\xmds\ permits the use of different spectral methods in each dimension.  Figure~\ref{fig:TensorProductLattice} is an example of a lattice using a Hermite-Gauss decomposition in the $x$ dimension and a Fourier decomposition in the $y$ dimension.  As discussed in Section~\ref{sec:GaussianQuadrature}, the grid spacing is determined by the choice of spectral method.  Full documentation of the spectral methods supported by \xmds\ and their uses is available from the \xmds\ website \citep{XMDSWebsite}.

\subsection{Fast spatial-to-spectral transforms}

In any nonlinear simulation, both the spatial and spectral representations of the solution will be required, as the problem will not be diagonal in either representation. Typically, the spatial representation is used for calculating the nonlinear terms, and the spectral representation for calculating derivatives.  The two are linked by a linear transformation which can in general be performed with a matrix multiplication.  The computational complexity of this operation is $\mathcal{O}(N^2)$ for a single dimension. In higher dimensions, the use of a tensor product lattice (see Figure~\ref{fig:TensorProductLattice}) enables the matrix multiplication to be factorised for each dimension.  In two dimensions for example, the computational complexity of a general spatial-to-spectral transformation is $\mathcal{O}(N_1^2 N_2 + N_1 N_2^2)$.  Without the use of a tensor product lattice, this cost would be $\mathcal{O}(N_1^2 N_2^2)$.

There are two cases in which we can reduce this computational cost: when we can use the Fast Fourier Transform (FFT) algorithm, or when the basis functions alternate in parity.

Spectral methods using complex exponentials, cosines or sines enable the use of the FFT algorithm and its variants for transformations between spatial and spectral representations.  These cost only $\mathcal{O}(N \log N)$ in one dimension or $\mathcal{O}(N_2 N_1 \log N_1 + N_1 N_2 \log N_2)$ in two dimensions.

If the basis functions have explicit, alternating parity $X_n(-x) = (-1)^n X_n(x)$ like the Hermite-Gauss functions, the Parity Matrix Multiplication Transform (PMMT) \citep{SpectralMethods} can be used which is faster than a direct matrix multiplication in each dimension. The idea is to separately transform the even and odd components of the solution, each of which costs $\mathcal{O}\left((N/2)^2\right)$ giving a total cost of $\mathcal{O}(N^2/2)$.  This factor of two reduction does not improve the overall scaling but can be a significant improvement for simulations dominated by the cost of the spatial-to-spectral transforms.

\subsection{Distributed memory parallelism}

\begin{figure}
    \centering
    \includegraphics[width=8cm]{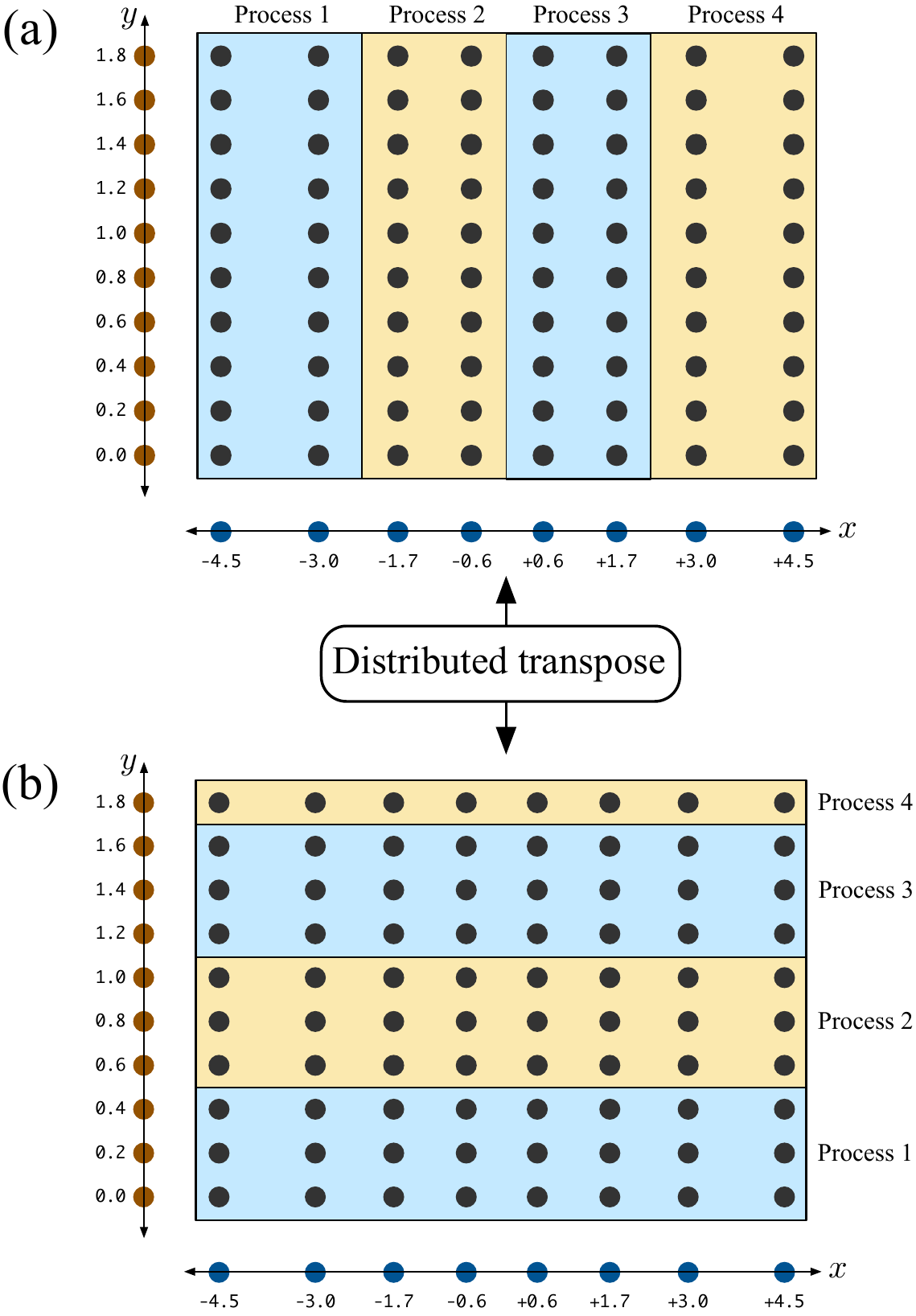}
    \caption{An example of problem parallelisation on a tensor product lattice. The problem is distributed across the (a) $x$ or (b) $y$ dimensions.  These two problem decompositions are linked by a distributed transpose operation.}
    \label{fig:TensorProductLatticeMPI}
\end{figure}

The use of a tensor product lattice permits the problem to be parallelised by distributing a single dimension across the available processes (see Figure~\ref{fig:TensorProductLatticeMPI}).  The advantage of this method is that as the spatial-to-spectral transform can be factorised across different dimensions, when the problem is decomposed across the $x$ dimension (as in Figure~\ref{fig:TensorProductLatticeMPI}(a)), the transform over the $y$ dimension can be performed as a purely \emph{local} operation to each process.  

To perform the spatial-to-spectral transform over the $x$ dimension, the problem must instead be decomposed across in the $y$ dimension (as in Figure~\ref{fig:TensorProductLatticeMPI}(b)).  As simulations typically require spatial-to-spectral transforms to be performed over all dimensions, a distributed transpose operation is used to link different problem decompositions (see Figure~\ref{fig:TensorProductLatticeMPI}).  This enables transforms to be performed over any dimension in a distributed simulation.

\subsection{Gaussian quadrature}
\label{sec:GaussianQuadrature}

Gaussian quadrature is an exponentially accurate method for integrating functions.  The key idea is to approximate
\begin{align}
    \int f(x)\, dx &\approx \sum_i f(x_i)\, w_i,
\end{align}
where $x_i$ are the interpolation points and $w_i$ are weight factors. Gaussian quadrature takes advantage of the fact that the interpolation points $x_i$ do not need to be equally spaced.  This means the $2N$ degrees of freedom $\{x_i, w_i\}$ can be chosen to exactly integrate $2N$ functions $f(x)$, while it would only be possible to exactly integrate $N$ functions if the $w_i$ were the only degrees of freedom. Further details about Gaussian quadrature are available from \citep{SpectralMethods,Stroud:1974}.

\subsection{Method-of-lines explicit temporal integration}

In method-of-lines integration, each grid point is considered to have its own ODE and the problem is integrated as a system of coupled ODEs.  \xmds\ employs a range of explicit integration schemes for deterministic and stochastic problems:
\begin{itemize}
    \item semi-implicit method (deterministic order 2, stochastic order 1) \citep{Werner:1997},
    \item fourth-order Runge-Kutta (deterministic order 4, stochastic order 1/2) \citep[\S 3.7(v)]{Abramowitz:1972},
    \item ninth-order Runge-Kutta (deterministic order 9, stochastic order 1/2) \citep{Tsitouras:2001},
    \item adaptive fourth-fifth order Runge-Kutta (deterministic only), \citep{Cash:1990} and
    \item adaptive eighth-ninth order Runge-Kutta (deterministic only) \citep{Tsitouras:2001}.
\end{itemize}

\xmds's fixed-step method-of-lines integration methods support integrating stochastic differential equations that depend on Wiener (Gaussian) or jump (counting) processes.  These stochastic differential equations must be entered in Stratonovich, not It{\^o} form \citep{GardinerHSM}.

Although the fourth-order Runge-Kutta and ninth-order Runge-Kutta algorithms have lower order stochastic convergence than the semi-implicit method, we find that they can be useful for problems where the noise terms are a perturbation on the `deterministic' dynamics.

\xmds\ can run multiple paths (possibly distributed across multiple processors) to compute moments of the stochastic process.  \xmds\ can also test the effect on the strong convergence \citep{KloedenPlaten} of the discretisation error of the propagation dimension.  This requires sampling the \emph{same stochastic trajectory} with timesteps of multiple sizes.

\subsection{Interaction picture method}

The method-of-lines integration schemes are supported by the interaction picture method \citep{Ballagh:1995,Caradoc-Davies:2000qy} which exactly solves a linear part of the differential equation.  

The idea is very similar to the interaction picture in quantum mechanics. The differential equation is split into two parts: a linear, exactly solvable component, and the remaining possibly nonlinear components.  The differential equation is then transformed to remove the exactly solvable component.  

For a PDE of the form
\begin{align}
    \frac{\partial f}{\partial t} &= \mathcal{L}[f] + g(x, y, f), \label{eq:PDEBeforeIPTransform}
\end{align}
where $\mathcal{L}$ is a linear operator that doesn't depend on time, the differential equation is transformed by defining the new quantity $\tilde{f} = e^{-\mathcal{L} t} f$ which evolves as
\begin{align}
    \frac{\partial \tilde{f}}{\partial t} &= e^{-\mathcal{L}t} g(x, y, e^{\mathcal{L} t} \tilde{f}). \label{eq:PDEAfterIPTransform}
\end{align}
The new quantity $\tilde{f}$ essentially has the simple dynamics due to $\mathcal{L}$ removed.

The interaction picture method is advantageous when the linear operator $\mathcal{L}$ has a faster characteristic timescale than the remainder of the differential system, which means that the function $\tilde{f}$ varies more slowly in time than the original $f$. This means that by solving the faster component separately and exactly, larger time-steps may be used on the remaining part of the differential equation while achieving the same solution accuracy.

For example, for the nonlinear Schr\"o{}dinger equation,
\begin{align}
    i \hbar \frac{\partial \psi}{\partial t} &= - \frac{\hbar^2}{2 m} \frac{\partial^2 \psi}{\partial x^2} + V(x) \psi + U \abs{\psi}^2 \psi, \label{eq:NonlinearSchrodingerEquation}
\end{align}
the spatial derivative term can have a faster characteristic timescale than the remainder of the system for high spatial resolutions, corresponding to high momentum components. In the Fourier basis, the spatial derivative term in Eq. (\ref{eq:NonlinearSchrodingerEquation}) becomes
\begin{align}
    \frac{\hbar^2 k_x^2}{2 m} \psi(k_x, t).
\end{align}
If we do not use the interaction picture, the maximum value of $k_x$ increases linearly with the number of grid points, and the time step used must decrease as $\Delta t \propto 1/N^2$ in order to be able to resolve the evolution of those terms.  The interaction picture method alleviates this problem by solving the spatial derivative term \emph{exactly}.  Using the interaction picture method to solve for the evolution of the spatial derivative term enables \eqref{eq:NonlinearSchrodingerEquation} to be solved with a time-step which is \emph{independent} of the spatial resolution.

\begin{figure}
    \centering
    \includegraphics[width=8cm]{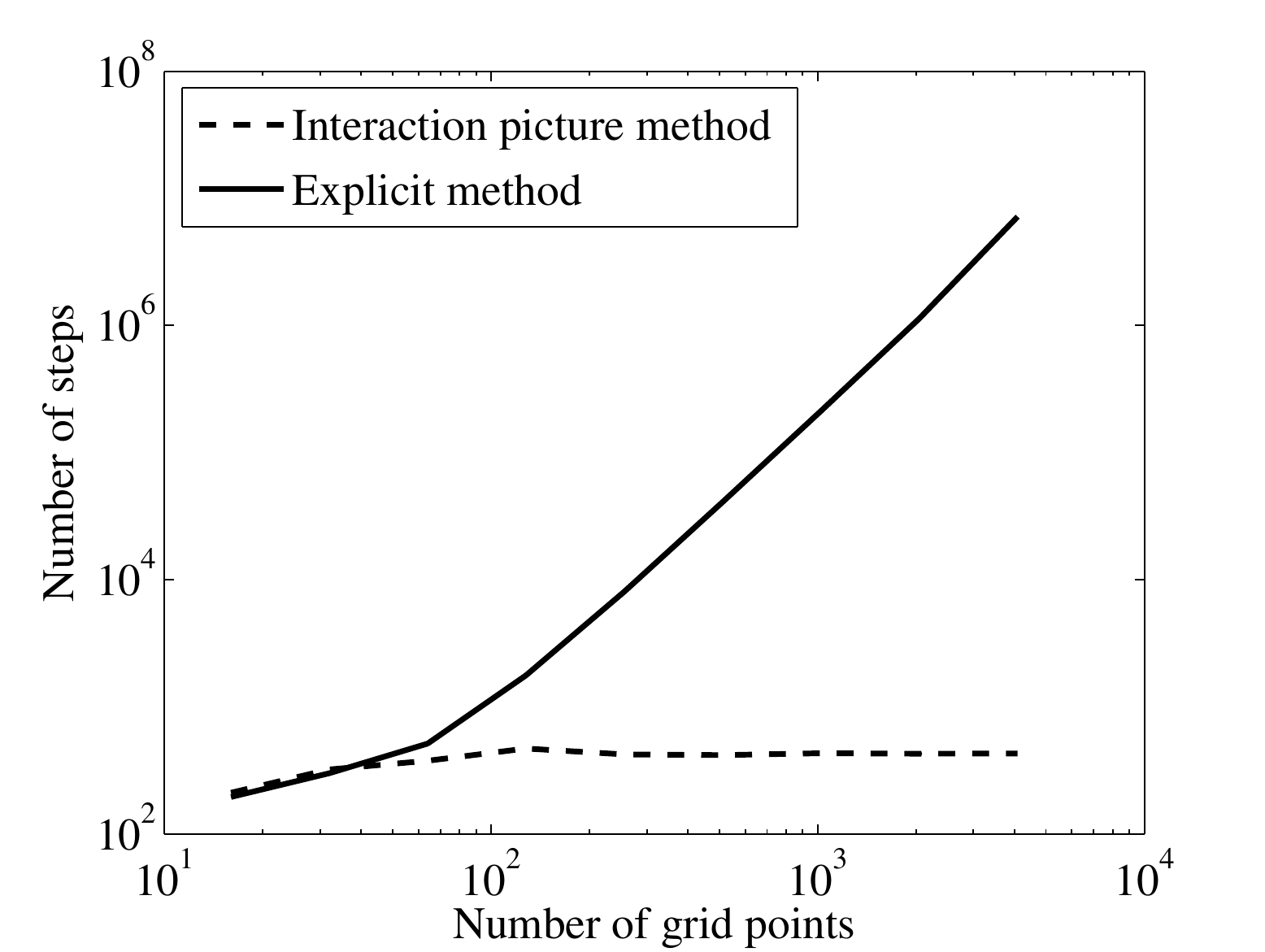}
    \caption{Comparison of the scaling of the interaction picture and `explicit' methods with grid resolution for computing the evolution due to spatial derivative terms.  Both methods were used to integrate the PDE \eqref{eq:NonlinearSchrodingerEquationDimensionless} with a fixed accuracy using an adaptive integrator.  As the resolution is increased, the number of steps remains approximately steady for the interaction picture method, while it increases quadratically (due to the second order spatial derivatives in \eqref{eq:NonlinearSchrodingerEquationDimensionless}) for the `explicit' method.  As the computational cost of each time step for both methods increases with resolution as $\mathcal{O}(N \log N)$ due to the use of Fourier transforms, the overall running time for the interaction picture method scales as $\mathcal{O}(N \log N)$ compared to $\mathcal{O}(N^3 \log N)$ for the `explicit' method. The XMDS2 scripts used can be found in \texttt{examples/cpc\_ip\_scaling.xmds} and \texttt{example/cpc\_ex\_scaling.xmds} in the XMDS2 distribution.}
    \label{fig:IPvsEX}
\end{figure}

The effect of the interaction picture method can be seen by solving \eqref{eq:NonlinearSchrodingerEquation} with an adaptive temporal integration method and comparing the number of time steps needed to achieve a given accuracy to that needed when calculating the derivatives explicitly (but still using a spectral method).  The results in Figure~\ref{fig:IPvsEX} demonstrate that the number of steps needed to solve the PDE using the interaction picture method is essentially independent of the spatial resolution, while for the explicit method, the number of steps needed increases quadratically.  

The computational cost of the interaction picture method is small if the linear operator $\mathcal{L}$ is local in either the spatial or spectral basis.  In this case, the application of $e^{\pm \mathcal{L} t}$ to the quantity $f$ can be calculated by transforming $f$ to the appropriate basis (spatial or spectral), performing a local multiplication, and transforming back to the original basis.  For fixed time-step algorithms, calculating the exponential function at every time step can be avoided by essentially redefining $\tilde{f}$ at each time step so that only the quantities $e^{\pm \mathcal{L} \Delta t}$ are needed.

To make best use of the interaction picture method, the basis functions should be chosen to make all derivative terms local in the spectral basis.  This ensures optimal scaling of the computational effort with spatial resolution.

Although the interaction picture method can be used with any integration software by applying the transformation manually, \xmds\ makes its use particularly easy by allowing the differential equation to be entered in a form equivalent to \eqref{eq:PDEBeforeIPTransform} with \xmds\ automatically making the transformation to \eqref{eq:PDEAfterIPTransform}.  This also enables easy comparisons to be made between the interaction picture and explicit methods.

\section{Examples}

In order to show the syntax of an \xmds\ script, as well as to demonstrate the ease with which simulations can be extended, we consider the behaviour of a Bose-Einstein condensate (BEC) in a harmonic magnetic trap.

\subsection{Example 1: Nonlinear Schr\"{o}dinger equation (\texttt{examples/cpc\_example1.xmds})}

Under a semiclassical approximation, the dynamics of the BEC will be governed by the nonlinear Schr\"{o}dinger equation with a harmonic trapping potential. In dimensionless units this equation is written
\begin{align}
    i \frac{\partial \psi}{\partial \bar{t}} &= -\frac{1}{2} \frac{\partial^2 \psi}{\partial \bar{x}^2} + \frac{1}{2} \bar{x}^2 \psi + U \abs{\psi}^2 \psi, \label{eq:NonlinearSchrodingerEquationDimensionless}
\end{align}
where $\bar{x} = \sqrt{m \omega /\hbar} \,x$, $\bar{t} = \omega t$, $m$ is the atomic mass, $\omega$ is the trapping frequency and $U$ is the nonlinear energy in units of $\hbar \omega$. \xmds\ is capable of solving much more complicated (sets) of PDEs, but this serves as a illustrative example.

Our initial condition will specify the wavefunction at $t=0$, and we choose
\begin{align}
\psi(\bar{x},0) = \sqrt{N} \, \pi^{-1/4} \exp ({-\bar{x}^2/2})
\label{eq:InitialCondition}
\end{align}
which is the ground state solution to Eq.~(\ref{eq:NonlinearSchrodingerEquationDimensionless}) in the absence of the nonlinearity, normalized to $N$ atoms in total.

\begin{figure*}
    \centering
    \includegraphics[width=14cm]{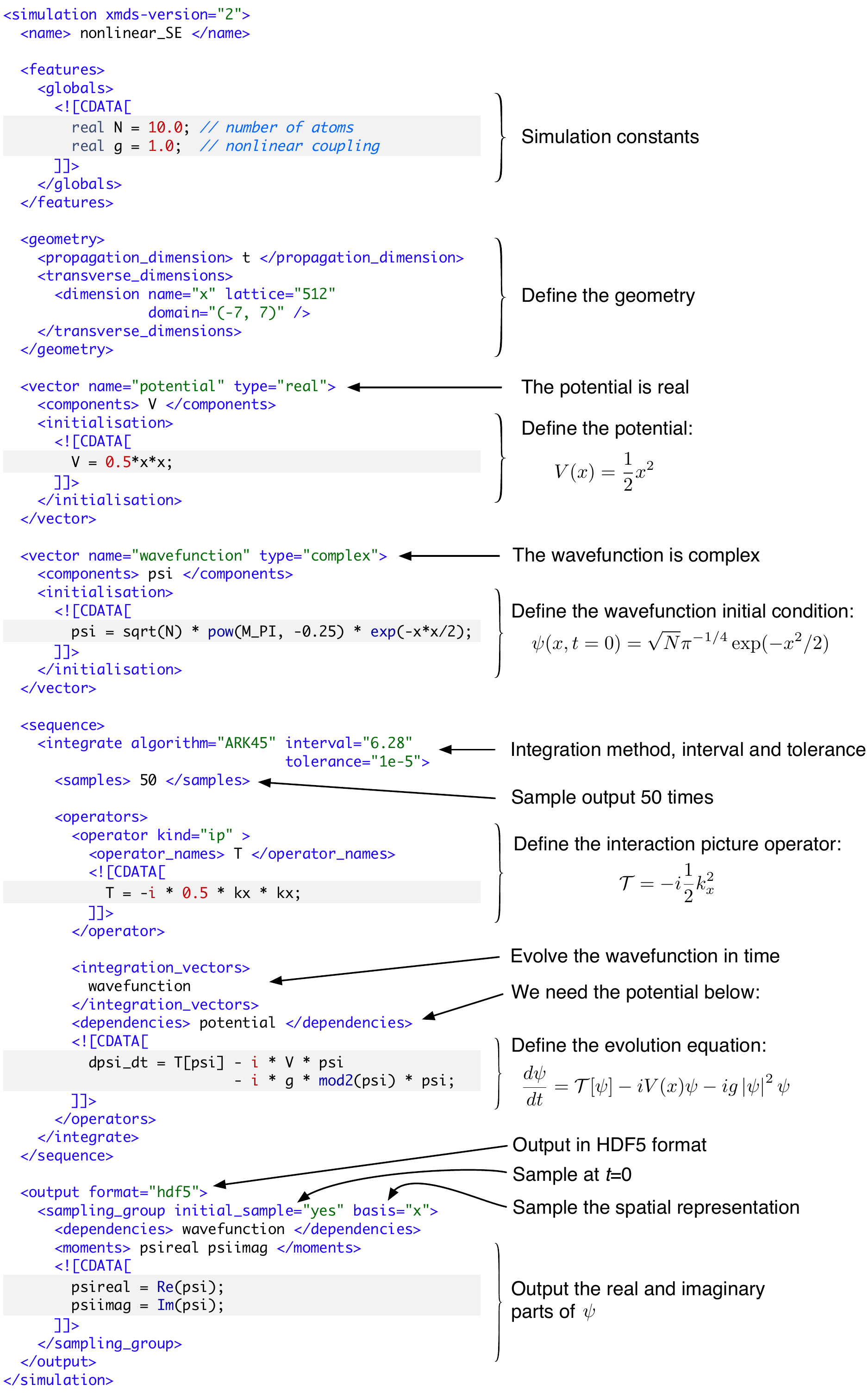}
    \caption{Annotated example \xmds\ script for integrating equation (\ref{eq:NonlinearSchrodingerEquationDimensionless}).  This script can be found in \texttt{examples/cpc\_example1.xmds} in the XMDS2 distribution.}
    \label{fig:examplescript}
\end{figure*}

We initially solve in one dimension, using a fourth-fifth order adaptive Runge-Kutta algorithm, evolving the system for a time $\bar{t} = 2 \pi /\omega$ (one trap period), sampling 50 times and outputting the real and imaginary parts of the wavefunction in position space at every grid point. An \xmds\ script to solve this problem is shown in Figure~\ref{fig:examplescript}.

\begin{figure}
    \centering
    \includegraphics[width=8.5cm]{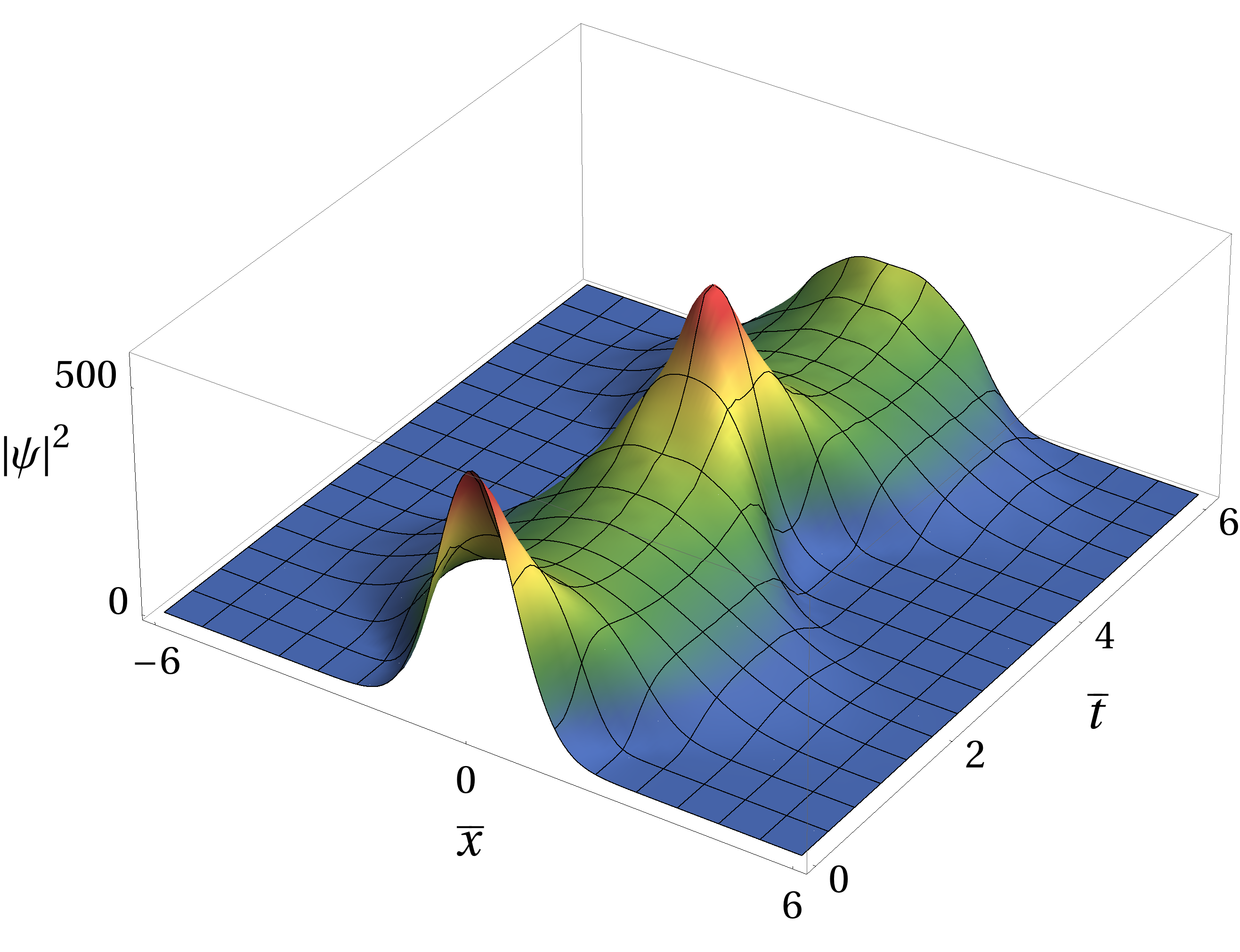}
    \caption{Solution to the nonlinear Schr\"{o}dinger equation given by Eq.~(\ref{eq:NonlinearSchrodingerEquationDimensionless}). The density $|\psi(x,t)|^2$ is shown evolving over one trap period.}
    \label{fig:nonlinear_breathing}
\end{figure}

When \xmds\ is run on this script, it produces an optimized binary {\small{\verb+nonlinear_SE+}} which is run to carry out the simulation. The result is shown in Figure \ref{fig:nonlinear_breathing}.

Changing parameters such as the domain or number of grid points, the number of sample points, integration interval, output moments, algorithm and precision used and so on is simply a matter of changing the contents of an XML tag, then re-running \xmds\ on the script to produce the new executable. While it is trivial to change such parameters, it is also easy to extend the simulation in more complex ways.

\subsection{Example 2: Higher dimensions (\texttt{examples/cpc\_example2.xmds})}

If one wished to run the simulation in two dimensions rather than one, all that is required is adding the element
{\small
\begin{verbatim}
<dimension name="y" lattice="512"
                    domain="(-7, 7)" />
\end{verbatim}
} 
\noindent
to the \small{\verb+transverse_dimensions+} element, changing the {\small{\verb+basis="x"+}} attribute of the {\small{\verb+sampling_group+}} element to \small{\verb+basis="x y"+}, adding a {\small{\verb+"0.5*y*y"+}} term to the potential {\small{\verb+"V"+}} and initial condition, and adding a {\small{\verb+"-i*0.5*ky*ky"+}} term to the kinetic energy operator {\small{\verb+"T"+}}.

\subsection{Example 3: Different transforms (\texttt{examples/cpc\_example3.xmds})}

This problem is obviously symmetric about $x=0$, so it is a waste of computational resources to simulate the problem on both sides of the origin. Since the differential equation and the boundary conditions are symmetric, by using the discrete cosine transform rather than the default exponential Fourier transform, we need only carry out the simulation on half the interval, and use only half the number of grid points for the same accuracy. This is accomplished simply by changing the content of the \small{\verb+transverse_dimensions+} element to be
{\small
\begin{verbatim}
<dimension name="x" lattice="256"
           domain="(0, 7)" transform = "dct"/>
\end{verbatim}
} 

\subsection{Example 4: Easy parallelization with MPI (\texttt{examples/cpc\_example4.xmds})}

\begin{figure}
    \centering
    \includegraphics[width=8cm]{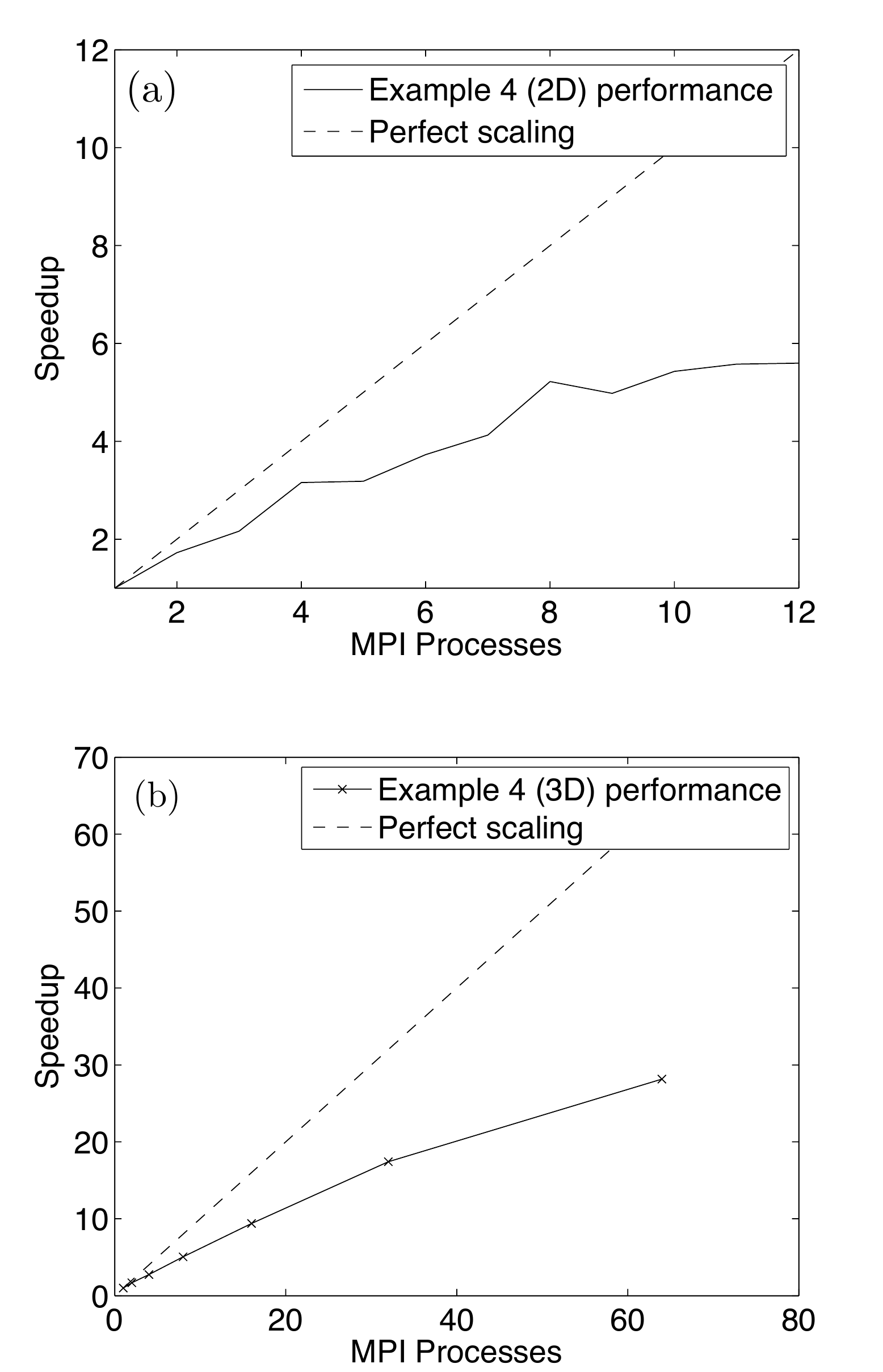}
    \caption{Example runtime scaling using MPI on (a) a single computer, and (b) a supercomputer. Figure (a) demonstrates the simulation \texttt{examples/cpc\_example4.xmds} run on a Linux computer with two Xeon 5675 CPUs running at 3.07GHz.  Each CPU has 6 execution cores.  Figure (b) demonstrates the simulation \texttt{examples/cpc\_example4\_3D.xmds} run on the NCI National Facility supercomputer, `vayu'.  The modified simulation is extended to three dimensions with 256 points in each to demonstrate performance on larger problems.  Note that optimal parallelisation for these problems is achieved when the number of grid points in the first dimension (256) is divisible by the number of processes.}
    \label{fig:parallelscaling}
\end{figure}

As this is a deterministic simulation, if it has two or more dimensions, one can parallelize the simulation simply by adding the {\small{\verb+<driver name="distributed-mpi" />+}} tag to the script. This would result in a binary that could be run across, for example, four CPUs with the command
{\small
\begin{verbatim}
mpirun -n 4 nonlinear_SE
\end{verbatim}
} 
The fact that two- or higher-dimensional deterministic simulations, as well as stochastic simulations of any dimension, can be trivially parallelized using a single line in a script, without spending days (or weeks) writing and debugging bespoke code, is one of \xmds's most powerful features.  The runtime scaling of this simulation with the number of processes is illustrated in Figure~\ref{fig:parallelscaling}.

\subsection{Example 5: Non-local terms (\texttt{examples/cpc\_example5.xmds})}

Many problems will involve non-local interactions that occur in the form of a convolution $\int f(r-r') g(r') \, dr'$. For example, within the context of the current problem, if the BEC were charged there would be an additional potential of the form
\begin{align}
V(x) = \frac{e^2 Z^2}{4 \pi \epsilon_0} \int \frac{1}{|x-x'| } |\psi(x')| ^2 \, d x'
\end{align}
where $eZ$ is the charge associated with each particle. While this term could be explicitly integrated within \xmds\, it is more efficient to make use of convolutions and the speed of fast Fourier transforms. This is done using \small{\verb+<computed_vector>+} elements, which are described in detail on our website \cite{XMDSWebsite}.

\subsection{Example 6: Stochastics (\texttt{examples/cpc\_example6.xmds})}

As a final tweak to this example, we will make use of \xmds's stochastic features to add noise. \xmds\ has a number of fast random number generators built in, which are capable of producing Gaussian, Poissonian and uniform probability distributions, and applying them as Wiener or jump processes during stochastic integration. This enables the simulation of stochastic differential equations, which are useful in fields such quantum field theory, mathematical finance, and many others. For this example we will simply use noise to model perturbations of the magnetic trap --- that is, the trapping potential will be slightly noisy, due to it moving around. To do this we define a noise vector
{\small
\begin{verbatim}
  <noise_vector name="trapNoise" kind="wiener"
                type="real" method="dsfmt">
    <components> noise_x </components>
  </noise_vector>
\end{verbatim}
} 
\noindent
change the potential term in the equation of motion in the script to
{\small
\begin{verbatim}
  - i * (V + g * mod2(psi) + alpha*noise_x) * psi
\end{verbatim}
} 
\noindent
where \small{\verb+alpha+} is a constant governing the magnitude of the noise, and add a \small{\verb+<dependencies>trapNoise</dependencies>+} tag to the \small{\verb+<initialisation>+} block of the \small{\verb+potential+} vector. This would add a time-dependent Gaussian-distributed noise to the potential. If we wished to average over many different realisations of this noise, we could add the tag
{\small
\begin{verbatim}
<driver name="mpi-multi-path" paths="100" />
\end{verbatim}
} 
\noindent
to the script, which would run the simulation 100 times, and average over whichever results were requested in {\small{\verb+<output>+}} section. Such a simulation could be trivially run over any number of CPU cores with near perfect scaling.  

Note that the {\small{\verb+mpi-multi-path+}} driver should only be used for stochastic simulations where individual realisations are independent, in contrast, the {\small{\verb+distributed-mpi+}} driver parallelises a single deterministic simulation.  As the different components of a deterministic simulation will in general be coupled, the {\small{\verb+distributed-mpi+}} driver necessarily incurs a larger communication overhead than the {\small{\verb+mpi-multi-path+}} driver.  In general, the {\small{\verb+distributed-mpi+}} driver can be used to parallelise a single realisation of a stochastic simulation, but if many paths are needed, the {\small{\verb+mpi-multi-path+}} driver will be preferable.

\section{Software used}

\xmds\ makes use of the following external libraries in its generated C++ simulations:
\begin{itemize}
    \item FFTW3 \citep{FFTW3} for FFTs and MPI distributed transpose operations,
    \item dSFMT \citep{Saito:2008} for random number generation,
    \item MPI for inter-process communication,
    \item HDF5 for data input and output, and
    \item GNU Scientific Library for special function evaluation.
\end{itemize}
\xmds\ itself also uses the following Python libraries when generating simulations: Cheetah, pyparsing, lxml, h5py, mpmath, and numpy.

\section{Conclusion}

Using \xmds\ for simulations accelerates development time, produces code that executes extremely quickly, and also produces a self-documenting workflow, as output data is wrapped with the compact XML code used to produce it.  

\xmds\ particularly excels at providing a smooth transition from a low-dimensional simulation to a higher-dimensional one, from a deterministic simulation to a stochastic one, or from a single-processor simulation to a distributed simulation running in parallel across multiple computers (or on a supercomputer). In hand-written codes, unless they were initially written with such a potential future extension in mind, each such change would require significant effort in rewriting the code. In \xmds\ such changes require only minimal change to the input script. This encourages users to create test simulations of a simpler system (e.g. reduced dimensionality), which makes the code run faster, allowing problems in the input script to be found and fixed more quickly. Later, the simulation can be scaled up to the full problem. Fundamentally, the ease with which codes can be generated encourages experimentation with different types of simulations, as the time taken to create the code is no longer the rate-limiting factor.

The installers, documentation and examples for \xmds\ can be found at the website \cite{XMDSWebsite}.  This same documentation is available in the \texttt{documentation/} directory of the XMDS2 distribution.

\section*{Acknowledgements}

We would like to thank B.\ Blakie for assistance with the Hermite-Gauss basis, M.\ Hush, R. Stevenson, and S. Szigeti for testing early versions of \xmds, and the \xmds\ community for testing and ideas.  We also acknowledge support from the NCI National Supercomputing Facility.

\bibliographystyle{elsarticle-num}
\bibliography{xmds}

\end{document}